\documentclass{eas}
\usepackage{graphicx}
\usepackage[round,colon,authoryear]{natbib}
\bibpunct{(}{)}{;}{a}{}{,}

\begin{document}

\title{Faint dwarf galaxies in nearby clusters} 
\author{I. Misgeld}\address{Universit\"ats-Sternwarte M\"unchen, Scheinerstr. 1, 81679 M\"unchen, Germany }
\author{M. Hilker}\address{European Southern Observatory, Karl-Schwarzschild-Strasse 2, 85748 Garching, Germany}
\author{S. Mieske}\address{European Southern Observatory, Alonso de Cordova 3107, Vitacura, Santiago, Chile}
\begin{abstract}
Besides giant elliptical galaxies, a number of low-mass stellar systems inhabit the cores of galaxy clusters, such as dwarf elliptical galaxies (dEs/dSphs), ultra-compact dwarf galaxies (UCDs), and globular clusters. The detailed morphological examination of faint dwarf galaxies has, until recently, been limited to the Local Group (LG) and the two very nearby galaxy clusters Virgo and Fornax. Here, we compare the structural parameters of a large number of dEs/dSphs in the more distant clusters Hydra\,I and Centaurus to other dynamically hot stellar systems.
\end{abstract}
\maketitle
\section{Dwarf galaxies and their star cluster mates: intriguing parallels}
 In \citet{2008A&A...486..697M, 2009A&A...496..683M}, we presented the analysis of the early-type dwarf galaxy population in the more distant clusters Hydra\,I and Centaurus, going beyond the well-studied Virgo and Fornax clusters. We investigated fundamental galaxy scaling relations, such as the colour--magnitude relation and the magnitude-surface brightness relation, as well as the faint end of the galaxy luminosity function, down to $M_V\sim -10$~mag, which is comparable to the LG dSph Sculptor.

\begin{figure}
	\centering
	\includegraphics[width=7.6cm]{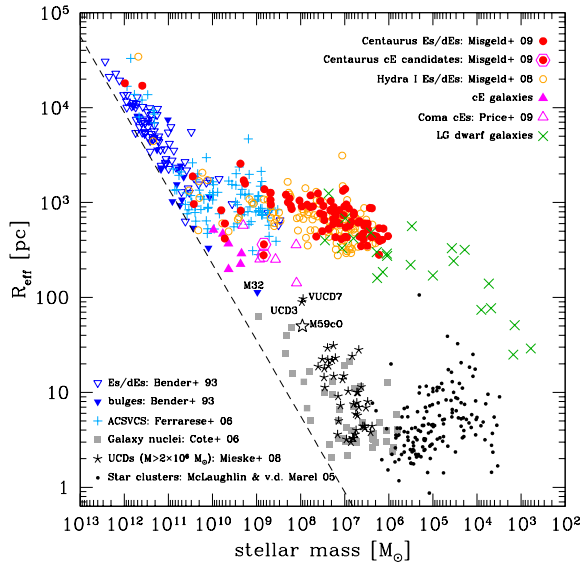}
	\caption{Effective radius $R_{\mathrm{eff}}$ plotted versus stellar mass $M_\star$ for various dynamically hot stellar systems. The dashed line indicates $R_{\mathrm{eff}} \propto M_{\star}^{4/5}$.}
	\label{fig:sizemass}
\end{figure}

Figure~\ref{fig:sizemass} shows the dwarf galaxies in the effective radius--stellar mass plane, in comparison to other dynamically hot stellar systems ranging from giant elliptical galaxies (gEs) with $10^{12}$~M$_\odot$ down to faint dwarf galaxies and star clusters of only a few $10^3$ solar masses. Two separate families of hot stellar systems can be identified: the 'galaxian' family, comprising gEs, Es, dEs and dSphs (coloured symbols), and the family of 'star cluster-like' objects, i.e. GCs, UCDs and nuclei of dE,Ns (black and grey symbols).

The sizes of dEs and dSphs ($R_{\mathrm{eff}} \sim 0.8$~kpc) on the one hand, and GCs ($R_{\mathrm{eff}} \sim 3$~pc) on the other hand, barely vary with mass over several orders of magnitude. However, the resulting size gap \citep[see also][]{2007ApJ...663..948G} gets filled in by low-mass, resolving star clusters and ultra-faint LG dwarf spheroidals at the low-mass end, and by GCs/UCDs, nuclei of dE,Ns and compact elliptical galaxies (cEs) in the mass range $10^6<M_\star<10^9$~M$_{\odot}$. The nearly constant mean effective radius of (dwarf) galaxies with masses $10^{11}<M_\star<10^6$~M$_{\odot}$ could serve as a distance indicator, provided it will be confirmed in more galaxy clusters.

Interestingly, massive ellipticals above $10^{11}$~M$_{\odot}$ show a similar size--mass relation as cEs, UCDs and nuclei of dE,Ns above $10^{6}$~M$_{\odot}$ \citep[see also][]{2008MNRAS.386..864D}. There is a clear common boundary towards minimum sizes, which can be approximated by $R_{\mathrm{eff}} \geq 2.24 \cdot 10^{-6} \cdot M_{\star}^{4/5}$~pc. No object, whether star cluster or galaxy, is located beyond this limit. If the size--mass relation is regarded as the consequence of a maximum possible stellar mass density, this might tell us something about how stars can be distributed in unrelaxed stellar systems.

\bibliographystyle{astron}
\bibliography{misgeld_dwarfs}

\end{document}